\documentstyle[aps,psfig]{revtex}
\newcommand{\HRule}{\rule{\linewidth}{1mm}}
\begin{document}
   \HRule
     \begin{center}
         \Huge {\bf ECT* PREPRINT}
     \end{center}
   \HRule
   \vspace*{\stretch{1}}
   \begin{center}
      \Large {\bf Superfluidity in $\beta$--stable neutron star matter} 
   \end{center}
   \begin{center}
      \large \O.\ Elgar\o y$^a$, L.\ Engvik$^a$, M.\ Hjorth--Jensen$^b$
        and E.\ Osnes$^a$
    \end{center}
    \begin{center}
     \large $^a$Department of Physics, University of Oslo, N-0316 Oslo,
     Norway
    \end{center}
    \begin{center}
      \large   $^b$ECT*, European Centre for Theoretical
        Studies in Nuclear Physics and Related Areas,
        Trento, Italy
    \end{center}
    \vspace*{\stretch{2}}
    \begin{center}
      \large   Submitted to: Physical Review Letters
    \end{center}
    \vspace*{\stretch{4}}
    \begin{center}
        \Large {\bf ECT* preprint $\#$: ECT*--96--009}
    \end{center}
    \begin{figure}[hbtp]
        \begin{center}
        {\centering\mbox{\psfig{figure=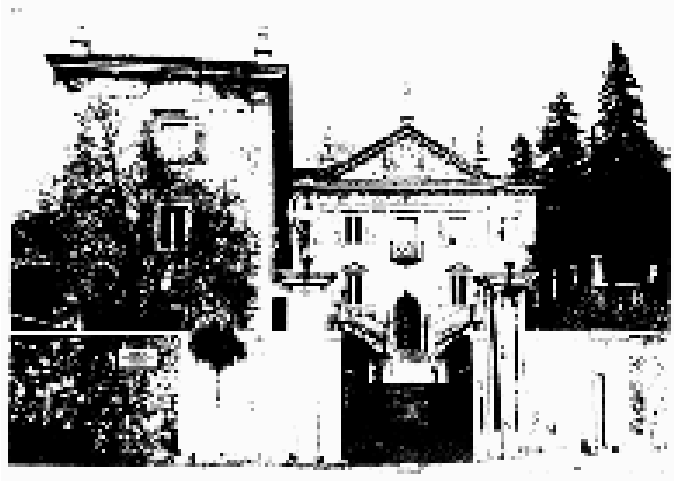,height=7cm,width=10cm}}}
        \end{center}
     \end{figure}
    \HRule
    \begin{center}
        \Large {\bf European Centre for Theoretical Studies in Nuclear
        Physics and Related Areas}
    \end{center}
    \begin{center}
        {\bf Strada delle Tabarelle 286, I--38050 Villazzano (TN),
        Italy}
    \end{center}
    \begin{center}
        {\bf tel.\ +39--461--314730, fax.\ +39--461--935007}
    \end{center}
     \begin{center}
        {\bf e--mail: ectstar@ect.unitn.it, www: http://www.ect.unitn.it}
    \end{center}
    \HRule

\clearpage

\draft

\title{Superfluidity in $\beta$--stable neutron star matter}
 
\author{\O.\ Elgar\o y$^a$, L.\ Engvik$^a$, M.\ Hjorth--Jensen$^b$
        and E.\ Osnes$^a$}

\address{$^a$Department of Physics, University of Oslo, N--0316 Oslo, Norway}

\address{$^b$ECT*, European Centre for Theoretical Studies in 
         Nuclear Physics and Related Areas, Trento, Italy}

\maketitle

\begin{abstract}

In this work we present results for pairing gaps in
$\beta$--stable neutron star matter with electrons and
muons
using a relativistic Dirac--Brueckner--Hartree--Fock
approach, starting with modern meson--exchange models for the 
nucleon--nucleon interaction.  
Results are given for superconducting $^1S_0$ protons
and $^3P_2$ and $^1D_2$ neutron superfluids. 
A comparison is made with recent non--relativistic calculations
and the implications for neutron star cooling are discussed.

\end{abstract}


Superfluidity and superconductivity of matter in neutron stars is 
expected to have a number of consequences directly related 
to observation, see Refs.\ \cite{st83,pr95,pethick92,page94,prakash94}. 
Among processes that will be affected are the
emission of neutrinos. Neutrino emission from e.g.\ 
various URCA processes  are expected to be the dominant 
cooling mechanism in neutron stars less than $10^5-10^6$ years old.
Typically,
proton superconductivity  reduces considerably the energy losses
in so--called modified URCA processes 
and may have important consequences for the
cooling of young neutron stars.
Another  possible manifestation of superfluid 
phenomena in neutron stars  
is glitches in rotational rates observed in a number
of pulsars. Moreover, 
the estimation of superfluid gaps and studies of pairing 
are not only important issues
in neutron star matter, but also in the rapidly developing 
field of neutron--rich systems such as heavy nuclei
close to the neutron drip line \cite{ms93} or the study of 
light halo nuclei \cite{riisager94}. 
Therefore, theoretical studies
of pairing in neutron--rich assemblies form currently a central
issue in nuclear physics and nuclear astrophysics.

The aim of this Letter  is to present results 
from self--consistent calculations for neutron and proton
pairing gaps in $\beta$--stable matter relevant for
neutron star studies. Pairing in 
the partial waves $^1S_0$, $^3P_2$ and $^1D_2$ will be studied using 
a relativistic approach with  
modern meson--exchange potential models to describe the nucleon--nucleon
(NN) potential.
A comparison with the corresponding non--relativistic approach is
also made. 

Our computational scheme is as follows:

The first ingredient in our calculation is the self--consistent
evaluation of single--particle energies in $\beta$--stable
matter starting from the meson--exchange potential models
of the Bonn group \cite{mac89}.  These single--particle
energies are obtained 
within the framework of the Dirac--Brueckner--Hartree--Fock (DBHF)
scheme \cite{cs86,bm90,hko95}, 
using a medium renormalized NN potential $G$ defined through
the solution of the $G$--matrix equation 
\begin{equation}
       G(\omega )=V+VQ\frac{1}{\omega - QH_0Q}QG(\omega ),
       \label{eq:bg}
\end{equation}
where $\omega$ is the unperturbed energy of the interacting  nucleons,
$V$ is the free NN potential, $H_0$ is the unperturbed energy of the
intermediate scattering states,
and $Q$ is the Pauli
operator which prevents scattering into occupied states.
Only ladder diagrams with two--particle states are included
in Eq.\ (\ref{eq:bg}).
In this work we solve Eq.\ (\ref{eq:bg}) using the Bonn A potential
defined in Table A.2 of Ref.\ \cite{mac89}. This potential model 
employs the Thompson \cite{bm90,thompson70} reduction of the
Bethe--Salpeter equation, and is tailored for relativistic
nuclear structure calculations. For further details, see 
Refs.\ \cite{mac89,bm90,hko95}.

The DBHF is a variational procedure where the single--particle
energies are obtained through an iterative self--consistency scheme.
To obtain the relativistic single--particle energies
we solve the Dirac equation for
a nucleon in the nuclear 
medium, with $c=\hbar=1$,
\begin{equation}
       (\not p -m +\Sigma (p))\tilde{u}(p,s)=0,
\end{equation}
where $m$ is the free nucleon mass and $\tilde{u}(p,s)$ is  
the Dirac spinor for positive energy solutions, with
 $p=(p^0 ,{\bf p})$ being
a four momentum and  $s$ the spin projection.
The self-energy $\Sigma (p)$ 
for nucleons can be written as
\begin{equation}
       \Sigma(p) =
       \Sigma_S(p) -\gamma_0 \Sigma^0(p)
       +\mbox{\boldmath $\gamma$}{\bf p}\Sigma^V(p).
\end{equation}
Since $\Sigma^V << 1$ \cite{bm90,sw86}, we approximate 
the self--energy by
\begin{equation}
       \Sigma \approx \Sigma_S -\gamma_0 \Sigma^0 = U_S + U_V,
\end{equation}
where $U_S$ is an attractive
scalar field and $U_V$ is the time-like component
of a repulsive vector field.
The Dirac spinor reads then
\begin{equation}
       \tilde{u}(p,s)=\sqrt{\frac{\tilde{E}_p+\tilde{m}}{2\tilde{m}}}
       \left(\begin{array}{c} \chi_s\\ \\
       \frac{\mbox{\boldmath $\sigma$}\cdot{\bf p}}
       {\tilde{E}_p+\tilde{m}}\chi_s
       \end{array}\right),
\end{equation}
where 
$\chi_s$ is the Pauli
spinor and 
terms with tilde like $\tilde{E}_p=\sqrt{{\bf p}^2+\tilde{m}^2}$
represent medium modified quantities.
Here we have defined \cite{bm90,sw86} $\tilde{m}=m+U_S$.
In all equations below, a momentum $p$ refers to the three--momentum
${\bf p}$.  
The single--particle energies 
$\tilde{\varepsilon}_p$ can then be written as
\begin{equation}
       \tilde{\varepsilon}_p=\tilde{E}_p +U_V.
       \label{eq:sprelen}
\end{equation}
The single--particle potential $u_p$ is given by 
$u_p=U_S\tilde{m}/\tilde{E}_p+U_V$ and 
can in turn be defined in terms
of the $G$--matrix
\begin{equation}
       u_p =\sum_{h\leq k_F} \frac{\tilde{m}^2}{\tilde{E}_h
       \tilde{E}_p}
       \left\langle  ph \right|
       G(\omega =\tilde{\varepsilon}_p
       +\tilde{\varepsilon}_h) \left| ph \right\rangle,
       \label{eq:urel}
\end{equation}
where $ph$ represent quantum numbers like momentum, spin, isospin projection
etc of the different single--particle states and $k_F$ is the Fermi
momentum. 
Eqs.\ (\ref{eq:sprelen})--(\ref{eq:urel}) are solved self--consistently
starting
with adequate values for the scalar and vector components
$U_S$ and $U_V$.
The proton fraction in $\beta$--equilibrium is
determined by imposing the relevant equilibrium conditions on the processes
$e^{-}+p \rightarrow n+\nu_{e}$
and $e^{-}\rightarrow \mu^{-}+\overline{\nu}_{\mu}+\nu_{e}$.
The conditions for $\beta$--equilibrium require that 
$\mu_{n}=\mu_{p}+\mu_{e}$,
where $\mu_{i}$ is the chemical potential of particle species $i$ ,
and that 
charge is conserved 
$n_{p}=n_{e}$,
where $n_{i}$ is the particle number density for 
particle species $i$.  We also include muons and the condition for charge 
conservation becomes 
$n_{p}=n_{e}+n_{\mu}$,
and chemical equilibrium gives
$\mu_{e}=\mu_{\mu}$.
Throughout we have assumed that neutrinos escape freely from the neutron  
star.  The proton and neutron chemical potentials are determined from 
the energy per baryon, calculated self--consistently in the above DBHF 
approach.

The next step in our calculations is to evaluate the pairing gaps for
various partial waves.
To evaluate the pairing gap we follow 
the scheme
of Baldo {\em et al.\ } \cite{bcll90}, originally
proposed by Anderson and Morel \cite{am61}. 
These authors introduced an
effective interaction $\tilde{V}_{k,k'}$.
This effective interaction
sums up all two--particle excitations 
above a cutoff momentum $k_M$, $k_M=3$ fm$^{-1}$ in this work. 
It is defined according to
\begin{equation}
       \tilde{V}_{k,k'}=V_{k,k'}-\sum_{k''>k_M}V_{k,k''}
       \frac{1}{2{\cal E}_{k''}}
       \tilde{V}_{k'',k'},
       \label{eq:gap1}
\end{equation}
where the energy ${\cal E}_k$ is given by
${\cal E}_k =\sqrt{\left(\tilde{\varepsilon}_k-
\tilde{\varepsilon}_F\right)^2+\Delta_k^2}$,
$\tilde{\varepsilon}_F$ being the single--particle energy at the Fermi surface,
$V_{k,k'}$ is the free nucleon--nucleon potential in momentum space, defined
by the three--momenta $k,k'$. 
The renormalized potential $\tilde{V}_{k,k'}$ and the free NN potential
$V_{k,k'}$ carry a factor $\tilde{m}^2/\tilde{E}_k\tilde{E}_{k'}$,
due to the normalization chosen for the Dirac spinors in nuclear matter.
These
constants are also included in the evaluation of the $G$--matrix,
as discussed in \cite{bm90,hko95}. 
For the $^1S_0$ channel, the pairing gap
$\Delta_k$ is \cite{kr90} 
\begin{equation}
       \Delta_k=-\sum_{k'\leq k_M}\tilde{V}_{k,k'}
       \frac{\Delta_{k'}}{2{\cal E}_{k'}}.
       \label{eq:gap3}
\end{equation}
For the $^3P_2$ partial wave we employ the expressions
given in Ref.\ \cite{eeho96a}, modified as well 
by the above normalization
constants. 
For further details, see e.g.\ Refs.\ \cite{bcll90,eeho96a,eeho96b}.
In summary, first we obtain the self--consistent 
DBHF single--particle spectrum $\tilde{\varepsilon}_k$ 
for protons and neutrons
in $\beta$--stable matter using the method
detailed in Ref.\ \cite{eeho96b}. 
Thereafter we solve self--consistently Eqs.\ (\ref{eq:gap1}) and
(\ref{eq:gap3}) in order to obtain the pairing gap
$\Delta$ for protons and neutrons for different partial waves.

\begin{figure}[htbp]
\begin{center}
{\centering
\mbox
{\psfig{figure=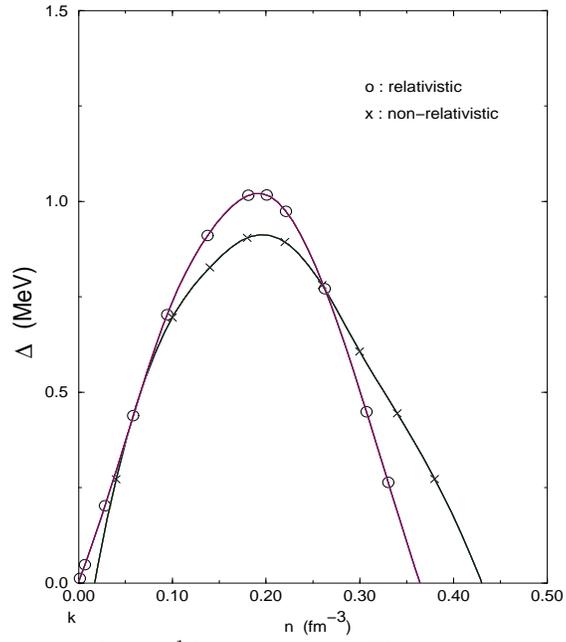,height=8cm,width=7cm}}
}
\caption{Proton pairing in $\beta$--stable matter for 
the $^1S_0$ partial wave. The non--relativistic results are taken from
Ref.\ [18].}
\label{fig:fig1}
\end{center}
\end{figure}

\begin{figure}[htbp]
\begin{center}
{\centering
\mbox
{\psfig{figure=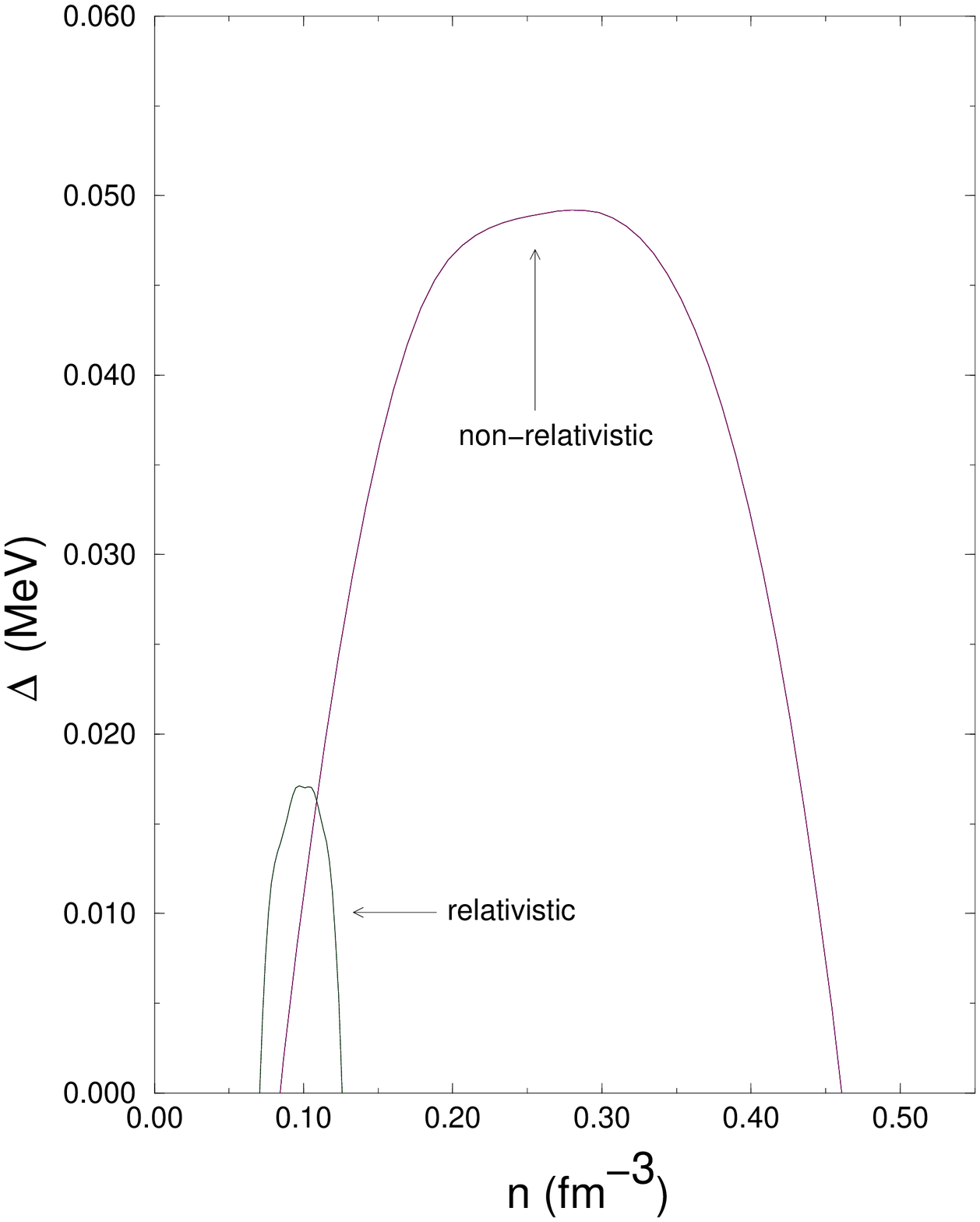,height=8cm,width=7cm}}
}
\caption{Neutron pairing in $\beta$--stable matter 
for the $^3P_2$
partial wave. The non--relativistic results are taken from Ref.\ [17].}
\label{fig:fig2}
\end{center}
\end{figure}

Our results for the pairing gaps, scalar and vector potentials
for neutrons and protons, proton and neutron fractions and the chemical
potential for electrons (and muons for total baryonic 
densities greater than $\rho=0.15$ fm$^{-3}$)
are displayed in Tables \ref{tab:tab1} and
\ref{tab:tab2} as functions of the total baryonic density.
The results  of these tables can in turn be used in relativistic equations
for various  modified URCA processes, in a similar way as done
in the non--relativistic approach of Friman and Maxwell \cite{fm79}. 
In Fig.\ \ref{fig:fig1} we plot as function of the total baryonic 
density the pairing gap for protons in the $^1S_0$
state, together with the results from the non--relativistic 
approach discussed in  Refs.\
\cite{eeho96b,eeho96c}. The results in the latter
references were  also obtained with the Bonn A potential of
Ref.\ \cite{mac89}. These results are all 
for matter in $\beta$--equilibrium. In Fig.\ \ref{fig:fig2} we plot the 
corresponding relativistic 
results for neutron energy gaps in the $^3P_2$ channel. For the 
$^1D_2$ channel we found both the non--relativistic and the relativistic
energy gaps to vanish. 
The non--relativistic
results for  the Bonn A potential are taken from Ref.\ \cite{eeho96a}. 
We have omitted a discussion on neutron pairing gaps in the
$^1S_0$ channel, since these appear at densities corresponding 
to the crust of the neutron star. The gap in the crustal material 
is unlikely
to have any significant effect on cooling processes \cite{pr95}.

As can be seen from Fig.\ \ref{fig:fig1}, there are only small differences
(except for higher densities)
between the non--relativistic and relativistic proton gaps in the 
$^1S_0$ wave. This is expected, since the proton fractions (and their respective
Fermi momenta) are rather
small, see Table \ref{tab:tab1}. 
For neutrons however, see Table \ref{tab:tab2},
the Fermi momenta are larger, and we would 
expect relativistic effects to be important. At Fermi momenta
which correspond to the
saturation point of nuclear matter, $k_F=1.36$ fm$^{-1}$
the lowest relativistic correction to the kinetic energy per 
particle is of the order of 2 MeV. At densities higher than the saturation
point, relativistic effects should be even more important, as can clearly
be seen in the calculations of Ref.\ \cite{bm90}. 
Since we are dealing with
very small proton fractions in Table \ref{tab:tab2}, 
a Fermi momentum
of $k_F=1.36$ fm$^{-1}$, would correspond to a total baryonic 
density $\sim 0.09$  fm$^{-3}$. Thus, at larger densities 
relativistic effects for neutrons should
be important.
This is also reflected in Fig.\ \ref{fig:fig2} for the pairing
gap in the $^3P_2$ channel
The relativistic $^3P_2$ gap is more than half
less than the corresponding non--relativistic one, and the 
density region is also much smaller. This is mainly due to the 
inclusion of relativistic single--particle energies in the 
energy denominator of Eq.\ (\ref{eq:gap3}) and the normalization
factors for the Dirac spinors in the NN potential.
Even the non--relativistic energy gaps for neutron star matter in
$\beta$--equilibrium are small compared with the results for pure
neutron
matter, where the $^3P_2$ energy gap 
has a maximum around $\sim 0.12-0.13$ MeV, see Refs.\
\cite{eeho96a,tt93,ao85}. 
The  consequences for cooling rates 
and the interior composition of a neutron
star are significant. A recent investigation of various cooling 
mechanisms by Schaab {\em et al.} \cite{swwg96} found that 
an agreement with observed surface temperatures 
was reached if the $^3P_2$
energy gaps were of the order $\sim 0.05$ MeV. Our non--relativistic
results for $\beta$--stable matter are of this size, while the 
relativistic energy gaps result in an almost  negligible 
suppression
of e.g.\ various modified URCA processes
in the interior of a neutron star.
These results, and those of Schaab {\em et al.} 
\cite{swwg96} as well,
are at askance with those of Page \cite{page94}, 
where, in order
to explain the observed temperature of Geminga, baryon pairing has to 
be present in most, if not all of the core of the star.

In summary, in this work we have calculated in a self--consistent
way single--particle energies and energy gaps using a relativistic
DBHF approach. To our knowledge, 
after the relativistic work of Kucharek and Ring \cite{kr90},
this is the first estimate of pairing gaps within the framework of the DBHF
approach. In Ref.\ \cite{kr90}, the $^1S_0$ gap in symmetric nuclear 
matter was studied
within the framework of the Serot--Walecka model \cite{sw86}. 
The only parameters which enter our approach are those which define the
free NN potential \cite{mac89}.
Here we have focused on  
pairing in dense matter, though our approach allows also for a consistent
treatment of other neutron star properties.
The same NN force used here has also been used in Ref.\ \cite{ehobo96}
to calculate the equation of state 
and the total mass and radius for a neutron star.
Combining the results from this work and those of Refs.\  
\cite{eeho96c,ehobo96},
the following picture emerges:\newline 
Within the DBHF approach,
the direct URCA processes are only allowed for densities larger 
than $0.52$ fm$^{-3}$, see Ref. \cite{eeho96c}. A neutron star
with total mass $1.6 M_{\odot}$ would have a central density of
$\rho_c=0.4$ fm$^{-3}$ in  $\beta$--stable matter \cite{ehobo96}.  
For such a central density, various modified URCA processes 
are possible  mechanisms for neutrino production in a neutron star.
The main suppression of these processes would then come
from protons in the $^1S_0$ state.  The reader should note that there
are other possible cooling mechanisms than those discussed here, 
such as neutrino--pair bremstrahlung \cite{pt94}, 
direct URCA with hyperons
or neutrino emissions from
more exotic states, such as pion and kaon condesates or quark matter,
see e.g.\ Refs.\ \cite{page94,prakash94,swwg96} for recent reviews.
However, for a star with central density $\rho_c=0.4$ fm$^{-3}$, 
many of these more exotic neutrino emissivities are less likely. 
Hyperons appear at densities $\rho \sim 0.3$ or greater \cite{kpe95}.
Similar densities are expected for kaons and quark matter 
\cite{swwg96,kpe95}. In addition, neutrino--pair bremstrahlung
was recently found \cite{pt94} to be much less important than previously
estimated. Thus, for a $1.6 M_{\odot}$ neutron star with central density of
$0.4$ fm$^{-3}$ obtained with our DBHF approach \cite{ehobo96}, the most likely
cooling scenario is through modified URCA processes, and the 
main suppression comes from superconducting protons in the $^1S_0$ state.
Finally, 
it ought to be noted that we have not included effects from 
medium polarization effects, as discussed in Ref.\ \cite{wap93}. These may
further change the size of the energy gaps and the neutrino
emissivities (the $^1S_0$ gap should  decrease
while the $^3P_2$ gap is expected to increase).
Further, we have not considered the possibility
of $^3D_2$ pairing, which appear due to the increased proton fraction,
as discussed by Alm {\em et al.\ } \cite{arsw96}.

This work has received support from The Research Council of Norway (NFR)
(Programme for Supercomputing) through a grant of computing time.
MHJ thanks the Istituto Trentino di Cultura, Italy,
and the NFR for financial support.

\begin{table}
\caption{Proton fractions $\chi_p$, scalar and vector
single--particle potentials
$U_S^p$ and $U_V^p$, respectively, for protons, 
the proton pairing gap $\Delta_p$ for protons in the
$^1S_0$ state and
the electron (and muon) chemical potential $\mu_e$
as functions of total baryonic density $\rho$. 
Densities are in units
of fm$^{-3}$ , $U_S^p$, $U_V^p$, 
$\Delta_p$ and $\mu_e$ in units of MeV.}
\begin{tabular}{rrrrrr}
\multicolumn{1}{c}{$\rho$}&
\multicolumn{1}{c}{$\chi_p$}&
\multicolumn{1}{c}{$U_S^p$}&
\multicolumn{1}{c}{$U_V^p$}&
\multicolumn{1}{c}{$\Delta_p$}&
\multicolumn{1}{c}{$\mu_e$}\\ \hline
   .0013 &     .0032 &  -7.8479 &   3.2471 &  .0121 & 11.7231 \\ 
   .0068 &     .0050 & -77.7002 &  61.7252 &  .0483 & 20.3904 \\
   .0281 &     .0096 & -172.0541&  135.3744&   .2024& 38.9884 \\
   .0583 &     .0156 &-236.5725 & 181.5207 &  .4386 & 58.5459 \\
   .0944 &     .0229 &-285.0128 & 213.1141 &  .7036 & 78.1881 \\
   .1377 &     .0307 &-329.1642 & 242.7944 &  .9107 & 98.3550 \\
   .1811 &     .0403 &-365.8355 & 270.4411 & 1.0160 & 115.8907 \\
   .2007 &     .0462 &-381.3338 & 283.5829 & 1.0173 & 123.0215 \\
   .2212 &     .0524 &-396.7707 & 297.3635 &  .9742 & 130.1985 \\
   .2627 &     .0658 &-424.5634 & 325.2710 &  .7712 & 143.9456 \\
   .3072 &     .0801 &-451.9637 & 357.1098 &  .4490 & 158.2441 \\
   .3304 &     .0877 &-464.7640 & 373.9551 &  .2638 & 165.5386 \\
   .3544 &     .0953 &-476.8407 & 391.2967 &  .1826 & 172.9228 \\
   .3594 &     .0968 &-479.2122 & 394.8924 &  .0856 & 174.4599 \\ 
\end{tabular}
\label{tab:tab1}
\end{table}

\begin{table}
\caption{Proton fraction $\chi_p$, neutron scalar and vector
single--particle potentials
$U_S^n$ and $U_V^n$, respectively and
the neutron  pairing gap $\Delta( ^3P_2)$
as functions of total baryonic density $\rho$. 
Densities are in units
of fm$^{-3}$ , $U_S^n$, $U_V^n$ and
$\Delta$ in units of MeV.}
\begin{tabular}{rrrrr}
\multicolumn{1}{c}{$\rho$}&
\multicolumn{1}{c}{$\chi_p$}&
\multicolumn{1}{c}{$U_S^n$}&
\multicolumn{1}{c}{$U_V^n$}&
\multicolumn{1}{c}{$\Delta( ^3P_2)$}\\ \hline
 .0756  & .0191 & -118.4076  &  90.1259 & 0.009 \\
 .0811  & .0202 & -127.8562  &  97.9057 & 0.013 \\
 .0849  & .0210 & -134.2159  & 103.1913 & 0.014 \\
 .0949  & .0230 & -150.7538  & 116.9925 & 0.017 \\
 .1012  & .0243 & -161.1272  & 125.6867 & 0.017 \\
 .1056  & .0252 & -167.9521  & 131.2468 & 0.017 \\
 .1125  & .0266 & -179.0345  & 140.6626 & 0.015 \\
 .1172  & .0275 & -186.6448  & 147.1867 & 0.013 \\
 .1196  & .0279 & -190.5106  & 150.5173 & 0.011 \\
\end{tabular}
\label{tab:tab2}
\end{table}

\end{document}